\documentclass{article}

\usepackage{subfigure}
\usepackage{graphicx}
\usepackage{xcolor}
\usepackage{geometry}
\usepackage{caption}
\usepackage{amsmath}
\usepackage[colorlinks=true,linkcolor=blue,citecolor=blue]{hyperref}
\usepackage{algorithm, algpseudocode}
\usepackage{comment}

\newtheorem{thm}{Theorem}[section]

\begin{document}
	
\title{An Algorithm for Generating Strongly Chordal Graphs}
\author{Md. Zamilur Rahman \\
School of Computer Science \\
University of Windsor \\
Windsor, Canada \\
\and Asish Mukhopadhyay \\
School of Computer Science \\
University of Windsor \\
Windsor, Canada \\
\and Yash P. Aneja \\
Odette School of Business \\
University of Windsor \\
Windsor, Canada}

\date{}
\maketitle{}

\begin{abstract}
Strongly chordal graphs are a subclass of chordal graphs. The interest in this subclass stems from the fact that many problems
which are NP-complete for chordal graphs are solvable in polynomial time for this subclass. However, we are not aware of any algorithm that can generate instances of this class, often necessary for testing purposes.  
In this paper, we address this issue. Our algorithm first generates chordal graphs, using an available algorithm 
and then adds enough edges to make it strongly chordal, unless it is already so. 
The edge additions are based on a totally balanced matrix characterizations of strongly chordal graphs.
\end{abstract}

\section{Introduction}
Let $G = (V,E)$ be an undirected graph with $n$ vertices in its vertex set $V$ and 
$m$ edges in its edge set $E$. 
A graph $G$ is strongly chordal if it is chordal and every even cycle of length 6 or more has a strong chord. 
Since the strongly chordal graphs are a subclass of the chordal graph, among 
many definitions of a strongly chordal graph this has a more intuitive connection 
with the parent class of chordal graphs, which have no induced cycle of size greater than 3. There are several NP-complete 
problems such as INDEPENDENT SET, CLIQUE, COLORING, CLIQUE COVER, DOMINATING SET, and STEINER TREE etc. can be solved 
efficiently for strongly chordal graphs. For example, the $k$-tuple domination problem in strongly chordal graph can be solved 
in linear-time if a strong ordering is provided~\cite{DBLP:journals/ipl/LiaoC03}. For testing purposes, we often need to 
generate instances of strongly chordal graphs. While there are algorithms to recognize 
strongly chordal graphs, we did not find any that can generate these. In this paper we address this issue. 


The following section introduces some notations and definitions. In section~\ref{cg}, we review an algorithm 
due to \cite{DBLP:journals/anor/MarkenzonVA08} for generating chordal graphs. In section~\ref{scg} we list some of the 
characterizations of strongly chordal graphs and in section~\ref{algorithm} we present our algorithm for the generation of 
strongly chordal graphs, establish its correctness and analyze its time complexity. Finally, section~\ref{conclusion} ends 
with some concluding remarks and open problems.

\section{Preliminaries}\label{preli}
A path in $G$ is a sequence of vertices $v_i, v_{i+1},\dots{.}, v_k$, where $\{v_j, v_{j+1}\}$ for $j=i, i + 1,\dots{.},k-1$, 
is an edge of $G$. A cycle is a closed path. The size of a cycle is the number of edges in it. A subset $S$ of $V$ is a clique 
if the induced subgraph $G[S]$ is complete. 

\begin{figure}[htb]
	\centering
	\subfigure[Chordal graph but not strongly chordal graph\label{Fig-CGExample}]{\includegraphics[scale=.5]{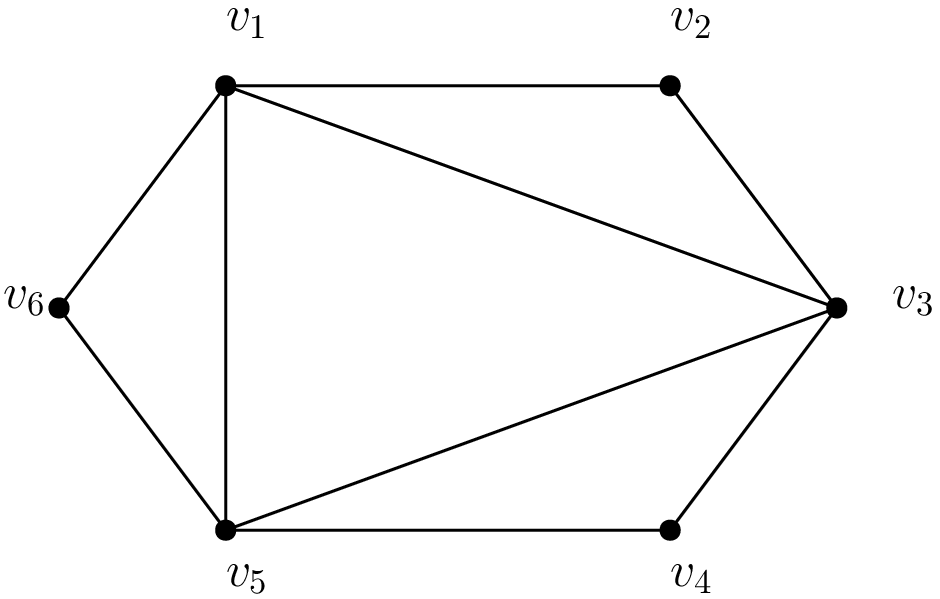}}\hspace{50pt}
	\subfigure[Strongly chordal graph \label{Fig-SCGExample}]{\includegraphics[scale=.5]{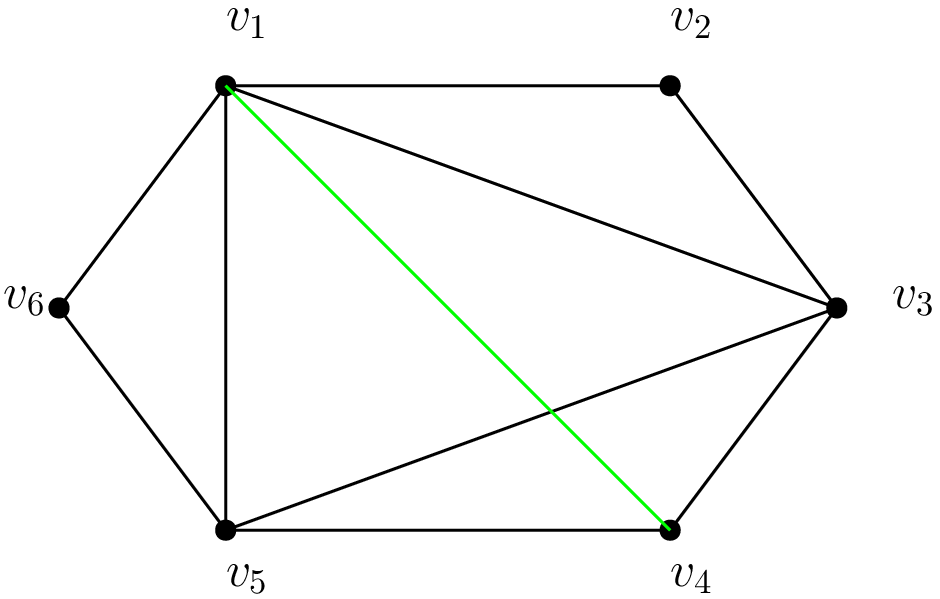}}
	\caption{An example}%
	\label{Fig:Examples}%
\end{figure}

A chord of a cycle is an edge joining two non-consecutive vertices. For instance, in Figure~\ref{Fig-CGExample}, the edge 
between $v_1$ and $v_3$ is a chord. A graph $G$ is said to be chordal if it has no chordless cycles of size $4$ or more. For 
example, both the graphs in Figure~\ref{Fig:Examples} are chordal.

A graph $G$ is strongly chordal if it is chordal and every even cycle of length $6$ or more has a strong chord. 
Figure \ref{Fig:Examples} shows an example of a chordal graph and a strongly chordal graph. The graph 
in Figure~\ref{Fig-CGExample}
 is chordal but not strongly chordal because there is no strong chord 
(in the literature this graph is known as the Hajos graph). On 
the other hand, the graph in Figure~\ref{Fig-SCGExample} is strongly chordal with $\{v_1, v_4\}$ (shown in green) as a strong chord.

Let $N(v)$ denote the neighborhood of a vertex $v$. The closed neighborhood of a vertex $v$, denoted by $N[v]$ is $N(v)\cup \{v\}$. The degree of a vertex $v$ is $\deg(v)=|N(v)|$. 

Let $G$ be a chordal graph. A vertex $v$ is said to be simplicial if $N(v)$ is a clique in $G$. A simplicial (perfect elimination) ordering of the vertices of $G$ is a map $\alpha:V\rightarrow\{1,2,\dots{.},n\}$ such that $v_i$ is simplicial in the induced graph on the vertex set $\{v_i, v_{i+1},\dots{.},v_n\}$. A graph $G$ is chordal if and only if there exists a perfect elimination ordering of its vertices. Thus  $v_2$, $v_6$, $v_4$, $v_1$ $v_3$, $v_5$ is a perfect elimination ordering of the
vertices of the chordal graphs shown in Figure~\ref{Fig:Examples}.

We have analogous definitions for a strongly chordal graph $G$. A vertex $v$ of $G$ is said to be simple if the sets in  $\{N[u]: u\in N[v]\}$ can be linearly 
ordered by inclusion. An alternate definition is this. Vertices $u$ and $v$ of $G$ are said to be compatible if 
$N[u]\subseteq N[v]$ or $N[v]\subseteq N[u]$ ~\cite{DBLP:journals/dm/Farber83}. 
Then a vertex $v$ is simple if the vertices in $N[v]$ are pairwise compatible. 
None of the vertices of the graph in  Figure~\ref{Fig-CGExample} is simple. Thus $v_2$ is not simple as the vertices  
$v_1$ and $v_3$ in $N[v_2] = \{v_1, v_2, v_3\}$ are not pairwise compatible.

A strong elimination ordering of a graph $G=(V,E)$ is an ordering $[v_1, v_2,\dots{.}, v_n]$ of $V$ such that the following condition holds: for each $i$, $j$, $k$, and $l$, if $i<j$, $k<l$ and $v_k, v_l\in N[v_i]$, and $v_k\in N[v_j]$, then $v_l\in N[v_j]$~\cite{DBLP:journals/dm/Farber83}. Thus a graph $G$ is strongly chordal if it admits a strong elimination ordering, which is a generalization of the 
notion of perfect elimination ordering used to define chordal graphs.  Since the graph shown in Figure~\ref{Fig-CGExample} is not a strongly chordal graph, there is no strong elimination ordering available. On the oher hand,  $v_2$, $v_6$, $v_4$, $v_1$ $v_3$, $v_5$ is a strong elimination ordering of the vertices of the graph shown in Figure~\ref{Fig-SCGExample}.

For non-adjacent vertices $u$ and $v$, a proper subset $S$ of $V$ is an $u-v$ separator if $u$ and $v$ lie in separate components of $G - S$.  
It is a minimal $u-v$ separator if no proper subset of $S$ is an $u-v$ separator. $S$ is a separator in $G$ if there exists 
vertices $u$ and $v$ in $V$ that $S$ separates. 

\section{Generation of chordal graphs}\label{cg}
 In~\cite{DBLP:journals/anor/MarkenzonVA08}, Markenzon et al. proposed two methods for the generation of chordal graphs. The first method adds edges incrementally, while maintaining chordality. The method is simple, dispensing with the need for any auxiliary data structure. The second method adds vertices incrementally, while maintaining a perfect elimination ordering of the vertices 
and also a clique-tree representation of the graph.  The first method generates sparse graphs, while the second one generates dense ones. 

We discuss the details of the first method. It makes crucial use of the following theorem. 

\begin{thm}\label{cgtheorem}
	\cite{DBLP:journals/anor/MarkenzonVA08}Let $G=(V,E)$ be a connected chordal graph and $u, v$ non-adjacent vertices of $V$. The augmented graph $G+(u,v)$ is chordal if and only if $G[V-I_{u,v}]$ is not connected where $I_{u,v} = N(u)\cap N(v)$.
\end{thm}

A formal description of the algorithm~\ref{algocg} based on theorem~\ref{cgtheorem} in given below.

\begin{algorithm}[H]
	\caption{ChordalGraphGeneration(G)}\label{algocg}
	\begin{algorithmic}[1]
		\State Generate a tree, $G$
		
		\State Pick two random vertices $u$ and $v$ from $G$.\label{pick}
		
		\If {there is an edge exists between these two vertices ($u$ and $v$)}
			\State go back to step~\ref{pick}.
		\Else
			\State Insert an edge between $u$ and $v$ by maintain the chordality property
		\EndIf
		
		\If {there is no common neightbors between $u$ and $v$}
			\State go back to step~\ref{pick}.
		\Else
			\State choose one vertex $x$ from the common neighbor of $u$ and $v$
		\EndIf
		
		\State Perform BFS from $u$ to $v$ on the graph consisting of all the adjacent vertices of $x$ minus the intersection of $u$ and $v$'s neighbors
		
		\If {BFS finds a path from $u$ to $v$} 
			\State go back to step~\ref{pick}.
		\Else
			\State insert an edge between $u$ and $v$
		\EndIf
	\end{algorithmic}
\end{algorithm}

The number of vertices, $n$, and the number of edges, $m$, are the two inputs to this method. This method starts with the generation of a tree with the given number of vertices $n$. During the generation of a tree, in every iteration either a new edge is inserted or an edge split into two edges, until the number of tree edges equals $n-1$. After the tree generation phase, $m-n+1$ more edges are added. The algorithm adds a new edge, provided chordality is preserved.

After the generation of a chordal graph, a perfect elimination ordering for this graph is computed based on the algorithm, named LEX-BFS proposed in~\cite{DBLP:journals/siamcomp/RoseTL76} by Rose et. al. The LEX-BFS algorithm is given below:
\begin{algorithm}[htb]
	\caption{LEX-BFS(G)}\label{algoLexBFS}
	\begin{algorithmic}[1]
		\State Assign the empty label list, (), to each vertex in $V$
		
		\For{$i \leftarrow n$ to $1$}
			\State Pick a vertex $v\in V$ with the lexicographically largest label list
			\State Set $\alpha(v) = i$
			\State For each unnumbered vertex $w$ adjacent to $v$, add $i$ to the label list of $w$
		\EndFor
		
		\State return $\alpha$
	\end{algorithmic}
\end{algorithm}

This perfect elimination ordering is used in the generation of strongly chordal graphs. After applying algorithm~\ref{algoscg}, 
the chordal graphs turns into strongly chordal graphs and the perfect elimination ordering also turns into a strong elimination 
ordering.

\section{Characterizations of Strongly Chordal Graphs}\label{scg}

Many different characterizations of strongly chordal graphs are extant in the literature. To make the paper self-contained, we state those
that are relevant to the problem at hand. Interestingly enough, for each 
chracterization of a chordal graph there seems to be a corresponding characterization of a strongly chordal graph. To drive home this, 
we have stated 
the first three characterizations of strongly chordal graphs along with their chordal counterpart.

The first characterization is based on elimination ordering. 

\begin{thm}\label{cgpeo}
	\cite{ROSE1970597}A graph is chordal if and only if it admits a perfect elimination ordering.
\end{thm}
\begin{thm}\label{scgseo}
	\cite{DBLP:journals/dm/Farber83}A graph is strongly chordal if it admits a strong elimination ordering.
\end{thm}

The second is based on the type of a vertex. 

\begin{thm}\label{cgsimplicial}
	\cite{ROSE1970597}A graph $G$ is chordal if and only if every induced subgraph $G$ has a simplicial vertex.
\end{thm}
\begin{thm}\label{scgsimple}
	\cite{DBLP:journals/dm/Farber83}A graph $G$ is strongly chordal if and only if every induced subgraph of $G$ has a simple vertex.
\end{thm}

The third pair is based on the absence of induced chordless cycles. 

\begin{thm}\label{cgtriangle}
	\cite{Chang_Nemhauser_1984}A graph is chordal if and only if every cycle of length greater than $3$ has an induced $1$-chord triangle.
\end{thm}
\begin{thm}\label{scgriangle}
	\cite{DBLP:journals/dm/DahlhausMM98}A graph is strongly chordal if and only if it has no chordless cycle on four vertices and every cycle on at least five vertices has an induced $2$-chord triangle.
\end{thm}

The next pair characterizations are based on totally balanced matrices for strongly chordal graphs only. The first characterization stated in theorem~\ref{scgmg} is based on the neighborhood matrix $M(G)$ of $G$ and the second characterization stated in theorem~\ref{scgcg} is based on the clique matrix $C(G)$ of $G$.

\begin{thm}\label{scgmg}
	\cite{DBLP:journals/dm/Farber83}The graph $G$ is strongly chordal if and only if $M(G)$ is totally balanced.
\end{thm}
\begin{thm}\label{scgcg}
	\cite{DBLP:journals/dm/Farber83}The graph $G$ is strongly chordal if and only if $C(G)$ is totally balanced.
\end{thm}

The last characterization of a strongly chordal graph is this:
 
\begin{thm}\label{scgstrongchords}
	\cite{DBLP:journals/dm/Farber83}A graph $G$ is strongly chordal if and only if it is chordal and every even cycle of length at least $6$ in $G$ has a strong chord.
\end{thm}

\section{Algorithm for Generating Strongly Chordal Graphs}\label{algorithm}
This section describes an algorithm for generating strongly chordal graphs. This algorithm takes the number of vertices, $n$, and the number of edges, $m$, as input. Based on this input, a chordal graph is generated using the incremental method described in section~\ref{cg}. Next, the chordal graph $G$ is passed as an input to the algorithm~\ref{algoscg} for generating a strongly chordal graph by introducing some additional edges, if needed. The algorithm is based on one of the characterizations by Farber~\cite{DBLP:journals/dm/Farber83}, which we have mentioned in theorem~\ref{scgmg}. For convenience, we restate it here:

\begin{thm}~\cite{DBLP:journals/dm/Farber83}\label{theoF}
	The graph $G$ is strongly chordal if and only if $M(G)$ is totally balanced.
\end{thm}
where $M(G)$ refers to the $n\times n$ neighborhood matrix of a graph $G$ on the vertices $v_1, v_2,\dots, v_n$ whose $(i, j)$ entry is $1$ if $v_i\in N[v_j]$ and is $0$ otherwise. The ordering of the vertices $v_1, v_2,\dots, v_n$ of a graph $G$ is a strong elimination ordering if and only if the matrix

\begin{equation*}
\Delta =
\begin{bmatrix}
1 & 1 \\
1 & 0
\end{bmatrix}
\end{equation*}

\begin{algorithm}[H]
	\caption{Generation of Strongly Chordal Graphs}\label{algoscg}
	\begin{algorithmic}[1]
		\Require A chordal graph $G$
		\Ensure A strongly chordal graph $G$
		
		\State Generate the neighborhood matrix $M(G)$ of $G$ from a perfect elimination ordering of a chordal graph $G$
		\While {there is a $\bigl[ \begin{smallmatrix}1 & 1\\ 1 & 0\end{smallmatrix}\bigr]$ submatrix present in $M(G)$}
		\State  Change the entry $0$ to $1$ in the submatrix and add a new edge in $G$ that corresponds to this entry
		\EndWhile
	\end{algorithmic}
\end{algorithm}
is not a submatrix of the neighborhood matrix, $M(G)$. Algorithm~\ref{algoscg} searches for the occurrences of $\Delta$ 
in $M(G)$, adding new edges to the graph 
whenever the 0 entry of a $\Delta$-matrix is changed to a 1. The  iteration continues until there is no $\bigl[ \begin{smallmatrix}1 & 1\\ 1 & 0\end{smallmatrix}\bigr]$ submatrix in $M(G)$. After applying algorithm~\ref{algoscg}, the chordal graph turns into a strongly chordal graph and also the perfect elimination ordering turns into a strong elimination ordering. \\

For testing purposes, at the end of this process, a recognition algorithm~\ref{algorecog} due to Farber \cite{DBLP:journals/dm/Farber83} is applied to verify that the graph generated by the algorithm~\ref{algoscg} is strongly chordal. When this 
algorithm terminates successfully it also generates a strong elimination ordering. A formal description of the recognition algorithm is given below:
\begin{algorithm}[H]
	\caption{REcognitionAlgorithm~\cite{DBLP:journals/dm/Farber83}}\label{algorecog}
	\begin{algorithmic}[1]
		\Require A graph $G=(V,E)$
		\Ensure A strong elimination ordering
		\State{Set $n\leftarrow |V|$.}
		\State{Let $V_0=V$ and let $(V_0,<_0)$ be the partial ordering on $V_0$ in which $v<_0 u$ if and only if $v=u$. Let $V_1=V$, and set $i\leftarrow 1$.}\label{step2}
		\State{Let $G_i$ be the subgraph of $G$ induced by $V_i$. If $G_i$ has no simple vertex then output $G_i$ and stop. Otherwise, define an ordering on $V_i$ by $v<_i u$ if $v<_{i-1}u$ or $N_i[v]\subset N_i[u]$.}
		\State{Choose a vertex $v_i$ which is simple in $G_i$ and minimal in $V_i,<_i$. Let $V_{i+1}=V_i-\{v_i\}$. If $i=n$ then output the ordering $v_1, v_2,\dots{.},v_n$ of $V$ and stop. Otherwise, set $i\leftarrow i+1$ and go to step~\ref{step2}.}
	\end{algorithmic}
	\label{alg_1}
\end{algorithm}
\begin{figure}[htb]
	\centering
	\subfigure[Chordal graph but not strongly chordal graph \label{Figure_CGSCGExample-2-1}]{\includegraphics[scale=.5]{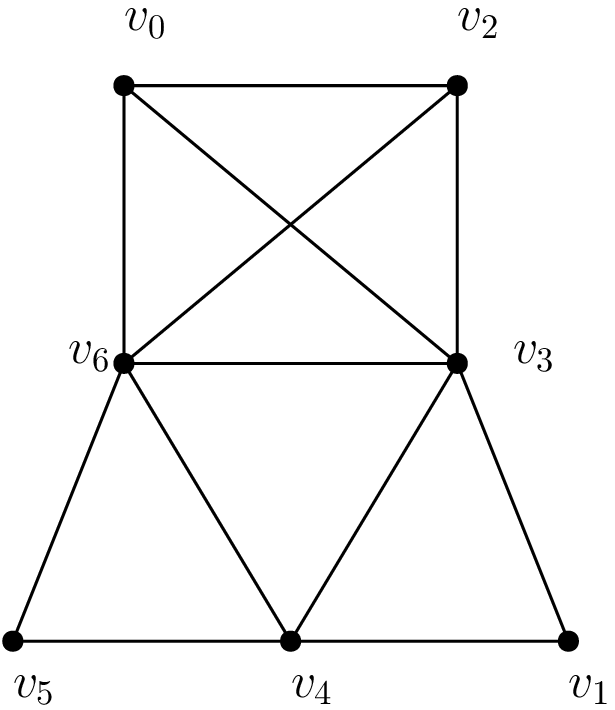}}\hspace{35pt}
	\subfigure[Strongly chordal graph\label{Figure_CGSCGExample-2-2}]{\includegraphics[scale=.5]{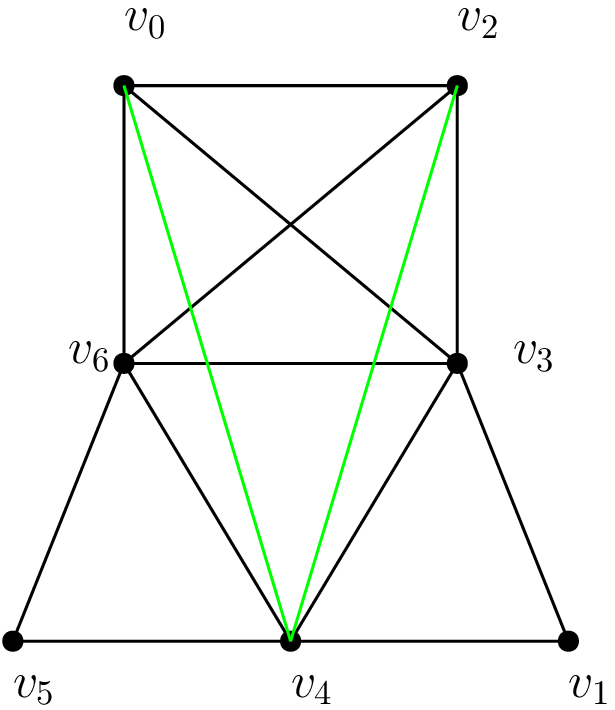}}\\
	\subfigure[Neighborhood matrix $M(G)$ of the chordal graph is shown in Figure (a)]
	{$\begin{bmatrix}
		1                    & 0                    & 0                    & 1   & 1   & 0   & 0   \\
		0                    & 1                    & 1                    & 1   & 0   & 0   & 1   \\
		0                    & 1                    & 1                    & 1   & 0   & 0   & 1   \\
		\textcolor{green}{1} & \textcolor{green}{1} & \textcolor{green}{1} & 1   & 1   & 0   & 1   \\
		\textcolor{green}{1} & \textcolor{green}{0} & \textcolor{green}{0} & 1   & 1   & 1   & 1   \\
		0                    & 0                    & 0                    & 0   & 1   & 1   & 1   \\
		0                    & 1                    & 1                    & 1   & 1   & 1   & 1
		\end{bmatrix}$}\hspace{35pt}
	\subfigure[Neighborhood matrix $M(G)$ of the strongly chordal graph is shown in Figure (b)]
	{$\begin{bmatrix}
		1                    & 0                    & 0                    & 1 & 1 & 0 & 0 \\
		0                    & 1                    & 1                    & 1 & 0 & 0 & 1 \\
		0                    & 1                    & 1                    & 1 & 0 & 0 & 1 \\
		\textcolor{green}{1} & \textcolor{green}{1} & \textcolor{green}{1} & 1 & 1 & 0 & 1 \\
		\textcolor{green}{1} & \textcolor{green}{1} & \textcolor{green}{1} & 1 & 1 & 1 & 1 \\
		0                    & 0                    & 0                    & 0 & 1 & 1 & 1 \\
		0                    & 1                    & 1                    & 1 & 1 & 1 & 1
		\end{bmatrix}$}
	\caption{An example of a chordal and a strongly chordal graph.}\label{Fig_Example-2}
\end{figure}

Figure~\ref{Figure_CGSCGExample-2-1} shows an example of a chordal graph and the strongly chordal graph generated by the 
above algorithm from this chordal graph. The graph shown in Figure~\ref{Figure_CGSCGExample-2-1} is chordal but not strongly chordal as there is no strong chord (no edge $\{v_0, v_4\}$ and $\{v_2,v_4\}$) in the six cycle $v_5$, $v_6$, $v_0$, $v_2$ $v_3$, $v_4$ and $v_1$, $v_3$, $v_2$, $v_0$ $v_6$, $v_4$. Also from the neighborhood matrix $M(G)$ it can be seen that there two 
$\bigl[ \begin{smallmatrix}1 & 1\\ 1 & 0\end{smallmatrix}\bigr]$ submatrices exist. According to the algorithm~\ref{algoscg}, by change those entires from $0$ to $1$, the submatrix will not exist anymore in the $M(G)$ or equivalently, by introducing two new edges $\{v_0, v_4\}$ and $\{v_2, v_4\}$. Now the graph also has strong chords in the six cycles $v_5$, $v_6$, $v_0$, $v_2$ $v_3$, $v_4$ and $v_1$, $v_3$, $v_2$, $v_0$ $v_6$, $v_4$. The resulting graph is now strongly chordal and the perfect elimination ordering $v_1$, $v_0$, $v_2$, $v_3$ $v_4$, $v_5$, $v_6$ turns into strong elimination ordering.
\begin{figure}[htb]
	\centering
	\subfigure[Chordal graph and also strongly chordal graph. \label{Figure_CGSCGExample-3-1}]{\includegraphics[scale=.65]{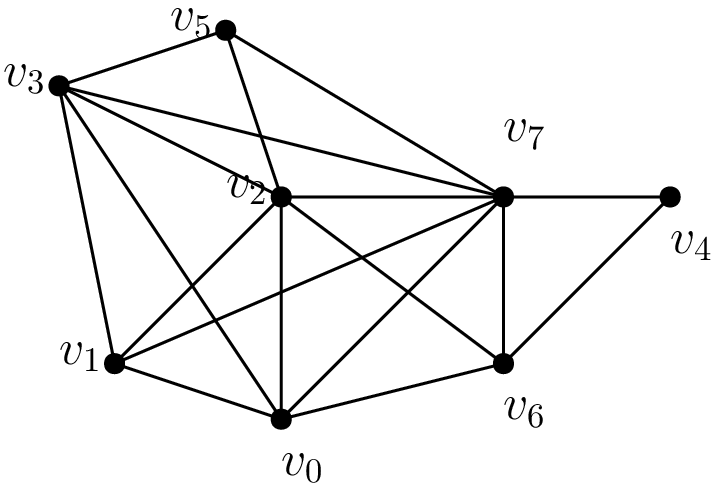}}
	\subfigure[Neighborhood matrix $M(G)$ of the strongly chordal graph is shown in Figure (a)]
	{$\begin{bmatrix}
		1   & 0   & 1   & 0   & 1   & 0   & 0   & 1   \\
		0   & 1   & 1   & 1   & 1   & 0   & 0   & 1   \\
		1   & 1   & 1   & 1   & 1   & 0   & 0   & 1   \\
		0   & 1   & 1   & 1   & 1   & 0   & 1   & 1   \\
		1   & 1   & 1   & 1   & 1   & 0   & 1   & 1   \\
		0   & 0   & 0   & 0   & 0   & 1   & 1   & 1   \\
		0   & 0   & 0   & 1   & 1   & 1   & 1   & 1   \\
		1   & 1   & 1   & 1   & 1   & 1   & 1   & 1
	\end{bmatrix}$}
	\caption{An example of a chordal and a strongly chordal graph.}\label{Fig_Example-3}
\end{figure}

Figure~\ref{Figure_CGSCGExample-3-1} shows another example of a chordal graph and the perfect elimination ordering is $v_5$, $v_1$, $v_3$, $v_0$ $v_2$, $v_4$, $v_6$, $v_7$. The graph is also strongly chordal because there are no $\bigl[ \begin{smallmatrix}1 & 1\\ 1 & 0\end{smallmatrix}\bigr]$ submatrices in the neighborhood matrix.
\begin{thm}
	Algorithm~\ref{algoscg} generates a strongly chordal graph, along with a strong elimination ordering.
\end{thm}
We know the ordering of the vertices $v_1, v_2,\dots, v_n$ of a graph $G$ is a strong elimination ordering if and only if the matrix
\[
\begin{bmatrix}
1 & 1 \\
1 & 0
\end{bmatrix}
\]is not a submatrix of the neighborhood matrix, $M(G)$ and the algorithm makes sure that no such submatrices are present in 
$M(G)$. Thus the resulting graph is a strongly chordal graph.


\subsection{Complexity}
A tree with $n-1$ edges is created for the given number of vertices $n$, as such a graph is chordal. Then an additional $m-n+1$ edges are added, manintaining chordality. The time complexity of this phase is $O(nm)$. LEX-BFS is a linear time algorithm on the given number of vertices and edges. Algorithm~\ref{algoscg} takes $O(n^4)$ time for finding the submatrices $\bigl[ \begin{smallmatrix}1 & 1\\ 1 & 0\end{smallmatrix}\bigr]$ and insertion of edges. After finding all such submatrices in one round and inserting edges, the algorithm makes further rounds until there is no such submatrix present in $M(G)$. In the worst
case, the initial graph is changed into a complete graph, which is strongly chordal. Thus the number of rounds is bounded above
by $O(n^2)$ and the comlexity of the entire algorithm is in $O(n^6)$.

\section{Discussion}\label{discussion}
\begin{figure}[htb]
	\centering
	\subfigure[Tree and also strongly chordal graph. \label{Fig-TreeExample}]{\includegraphics[scale=.65]{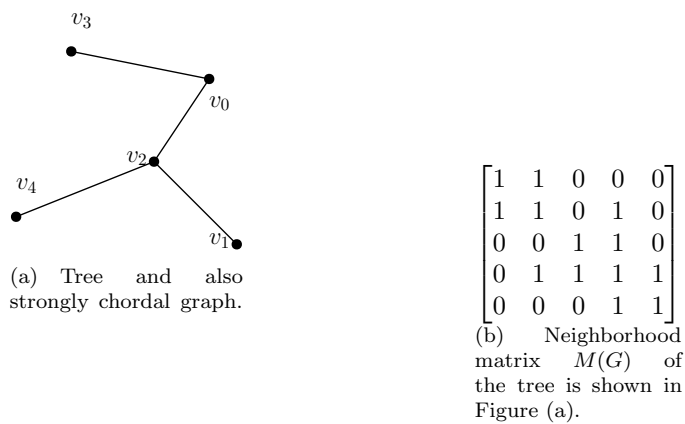}}\hfil
	\subfigure[Neighborhood matrix $M(G)$ of the tree is shown in Figure (a). \label{Fig-TreeMG}]
	{$\begin{bmatrix}
		1   & 1   & 0   & 0   & 0 \\
		1   & 1   & 0   & 1   & 0 \\
		0   & 0   & 1   & 1   & 0 \\
		0   & 1   & 1   & 1   & 1 \\
		0   & 0   & 0   & 1   & 1
		\end{bmatrix}$}
	\caption{An example of a tree with neighborhood matrix.}\label{Fig_Example-4}
\end{figure}

We proposed an algorithm to generate strongly chordal graphs on $n$ vertices and $m$ edges as input. We could skip the intermediate phase of generating chordal graphs and generate strongly chordal graphs directly from trees as these are chordal. However, this does not allow us to add new edges in the tree because a tree is also strongly chordal and there is no $\Delta$ submatrix present in the neighborhood matrix of a tree. Hence we get very sparse strongly chordal graphs, identical with the input trees. Figure~\ref{Fig-TreeExample} shows an example of a tree (also strongly chordal graph) and the perfect elimination ordering is $v_3$, $v_0$, $v_1$ $v_2$, $v_4$. The neighborhood matrix is given in Figure~\ref{Fig-TreeMG} from which we can see that there is no $\Delta$ submatrix present. An interesting open problem is to generate strongly chordal graphs ab initio, skipping the intermediate phase of generating chordal graphs. This might possibly 
lead to a more efficient algorithm than proposed in this paper. 

\end{document}